\documentclass[aps,prx,reprint,twocolumn,oneside,floatfix,sort&compress,superscriptaddress,showpacs,nobalancelastpage,longbibliography]{revtex4-2}
\usepackage{graphicx}
\usepackage{amssymb,bbm}
\usepackage[reqno]{amsmath}

\usepackage[colorlinks,bookmarks=false,citecolor=blue,linkcolor=red,urlcolor=blue,hyperfootnotes=true]{hyperref}

\usepackage{comment}

\newcommand{\be}{\begin{equation}}
\newcommand{\ee}{\end{equation}}
\newcommand{\bea}{\begin{eqnarray}}
\newcommand{\eea}{\end{eqnarray}}
\newcommand{\ban}{\begin{eqnarray*}}
\newcommand{\ean}{\end{eqnarray*}}
\newcommand{\dagga}{{\phantom{\dagger}}}
\newcommand{\bR}{\mathbf{R}}
\newcommand{\bQ}{\mathbf{Q}}

\newcommand{\bq}{\mathbf{q}}

\newcommand{\bk}{\mathbf{k}}

\newcommand{\br}{\mathbf{r}}
\newcommand{\ba}{\mathbf{a}}
\newcommand{\bb}{\mathbf{b}}

\newcommand{\bS}{\mathbf{S}}
\newcommand{\bnot}{\mathbf{0}}

\newcommand{\dis}{\displaystyle}

\newcommand{\fract}[2]{\frac{\dis #1}{\dis #2}}

\newcommand{\eqn}[1]{(\ref{#1})}

\newcommand{\BCS}{BaCoS$_2$}
\newenvironment{eqs}%
{\begin{equation} \begin{aligned}}%
{\end{aligned} \end{equation} }
\newcommand{\beal}{\begin{eqs}}
\newcommand{\eal}{\end{eqs}}
\newcommand{\bealn}{\beal\nonumber}
\newcommand{\bw}{\begin{widetext}}
\newcommand{\ew}{\end{widetext}}
\newcommand{\esp}[1]{\text{e}^{#1}}

\newcommand{\bd}[1]{{\boldsymbol{#1}}}
\begin{document}

\title{Order from disorder phenomena in BaCoS$_2$}

\author{Benjamin Lenz}
 \email{benjamin.lenz@upmc.fr}
 \affiliation{IMPMC, Sorbonne Universit\'e, CNRS, MNHN, 4 place Jussieu, 75005 Paris, France.}
 
 \author{Michele Fabrizio}
 \email{fabrizio@sissa.it}
 \affiliation{International School for Advanced Studies (SISSA), Via Bonomea 265, I-34136 Trieste, Italy}
 
\author{Michele Casula}
 \email{michele.casula@upmc.fr}
\affiliation{IMPMC, Sorbonne Universit\'e, CNRS, MNHN, 4 place Jussieu, 75005 Paris, France.}

\date{\today}                 

\begin{abstract}
At $T_N\simeq 305~\text{K}$ the layered insulator BaCoS$_2$ transitions to 
a columnar antiferromagnet that signals non-negligible magnetic frustration despite the relatively high $T_N$, all the more surprising given its quasi two-dimensional structure. Here, we show, by combining \textit{ab initio} and model calculations,  
that the magnetic transition is an order-from-disorder phenomenon, which 
not only drives the columnar $C_4\to C_2$ symmetry breaking, but also, and more importantly, the inter-layer coherence responsible 
for the finite N\'eel transition temperature. This uncommon ordering mechanism, actively contributed by orbital degrees of freedom, hints at an abundance of low energy excitations below and, especially, above $T_N$, not in disagreement with experimental evidences, and might as well emerge in other layered correlated compounds showing frustrated magnetism at low temperature.  
\end{abstract}

\maketitle


There is growing interest in the BaCo$_{1-x}$Ni$_x$S$_2$, $x\in[0,1]$, series 
for it displays a rich physics comprising renormalized Dirac states \cite{nilforoushan2021moving,PhysRevResearch.2.043397}, non-Fermi liquid behavior \cite{PhysRevB.93.125120,David2018}, and  insulator-to-metal transitions as  function of both doping and pressure \cite{Martinson1993,Sato1998,Kanada1999,Yasui1999,Zainullina2012,Guguchia2019}, typical manifestations of underlying strong electronic correlations. Here, we focus on the left-hand side of that series, i.e., $\text{Ba}\text{Co}\text{S}_2$, 
which undergoes at the N\'eel temperature 
$T_N\simeq 305~\text{K}$~\cite{Mandrus1997}, more recently estimated to be 
$T_N\simeq 290~\text{K}$~\cite{Abushammala2023}, 
a transition from a correlated paramagnetic insulator to 
an antiferromagnetic one, characterised by columnar spin-ordered planes, which we 
hereafter refer to as antiferromagnetic striped (AFS) order. The planes are 
stacked ferromagnetically along the $c$-axis, so called C-type stacking as opposed to the antiferromagnetic G-type one. \\
Inelastic neutron scattering experiments
show that magnetic excitations below $T_N$ have pronounced two-dimensional (2D) nature~\cite{Shamoto-JPSJ1997,Shamoto-PRR2021}. 
For instance, a direct estimate of spin exchange constants by neutron diffraction 
has been recently attempted 
in doped tetragonal BaCo$_{0.9}$Ni$_{0.1}$S$_{1.9}$
subject to an uniaxial strain~\cite{Shamoto-PRR2021}. 
This compound undergoes a N\'eel transition to the C-type AFS phase at 280~K~\cite{Shamoto-PRR2021}, not far from $T_N$ of undoped BaCoS$_2$. 
The spin model used to fit neutron data was a conventional 
$J_1-J_2$ Heisenberg model~\cite{J1-J2-original} on each plane, 
supplemented by an inter-plane ferromagnetic exchange $J_c$
estimated to be $\lesssim 0.04~\text{meV}$.
The latter is much too small to explain through the estimated 
$J_2\sim 9.3~\text{meV}$ and $J_1 \sim -2.3~\text{meV}$~\cite{Shamoto-PRR2021}
the large value of $T_N\simeq 290~\text{K}$, which is therefore puzzling and highly surprising.  \\ 
Another startling property is the anomalously broad peak of the magnetic susceptibility at $T_N$~\cite{Mandrus1997,Abushammala2023}, which hints at a transition in the Ising universality class rather than the expected Heisenberg one~\cite{Mandrus1997}. 
A possible reason of this behavior might be the spin-orbit coupling~\cite{Mandrus1997}. 
Indeed, a Rashba-like spin-orbit coupling due to the layered structure and the staggered sulfur pyramid orientation, see Fig.~\ref{fig:Mag}, has been found to yield sizeable band splittings at specific points within the Brillouin zone, at least in metallic BaNiS$_2$~\cite{Michele-SOC2016}. 
The Rashba-like spin-orbit coupling strength may barely differ in BaCoS$_2$, 
or be weakened by strong correlations~\cite{Capone-PRB2020}. 
In either case, its main effect is to introduce an easy plane anisotropy, as indeed  observed experimentally~\cite{Mandrus1997}, which, at most, drives the transition towards the $XY$ universality class. 
It is well possible that the weak orthorhombic distortion may turn the easy plane into an easy axis, but the resulting magnetic anisotropy should be negligibly small, being a higher-order effect, and thus unable to convincingly  
explain the experimental observations. \\
Our aim here is to shed light into 
those puzzles, which we achieve by a 
combination of \emph{ab initio} calculations and model 
analyses.\\

Before entering into the details of our work, it is worth anticipating the results and placing them in a more general framework. 
Frustrated magnets, as BaCoS$_2$ certainly is, often display a continuous accidental degeneracy of the classical ground state that leads 
to the existence of pseudo-Goldstone modes within the harmonic spin-wave approximation~\cite{Rau-PRL2018}. 
Since those modes are not protected by symmetry, they may acquire a mass once anharmonic terms 
are included in the spin-wave Hamiltonian. This mass, in turn, cuts off the 
singularities  
brought about by the pseudo-Goldstone modes, 
in that way stabilizing ordered phases otherwise thwarted by fluctuations.  
Put differently, 
let us imagine that the classical potential has a manifold of degenerate minima generally not invariant under the symmetry group of the Hamiltonian. 
It follows that the eigenvalues of the Hessian of the potential change from minimum to minimum. Allowing for quantum or thermal fluctuations is therefore expected to favour the minima with lowest eigenvalue, although the two kinds of fluctuations not necessarily select the same ones~\cite{Zhitomirsky-PRB2020}. Moreover, it is reasonable to assume that the minima with lowest eigenvalues are those that form subsets invariant under a symmetry transformation of the Hamiltonian, so that choosing any of them corresponds to a spontaneous symmetry breaking. Such a phenomenon,  
also known as order from disorder~\cite{Villain-1980}, emerges in many different contexts~\cite{Burgess_2000}, from particle physics~\cite{Weinberg-PRL1972,Weinberg-PRD1973} to condensed matter physics~\cite{Zhang-RMP2004,Chubukov-RPP2016}, even though frustrated magnets still provide the largest variety of physical realisations~\cite{Tessman-PR1954,Villain-1980,Belorizky-1980,shender-1982,Henley-PRL1989,J1-J2-original,Zhitomirsky-JETP2005,Rau-PRL2014,Rau-PRL2018, Zhitomirsky-PRB2020}.\\
Our results indicate that the antiferromagnetic insulator BaCoS$_2$ is another representative of the class of frustrated magnets where order from disorder effects play a very active role and may explain at once the puzzling phenomenology of this material, including very recent experimental findings, presented and discussed in a parallel publication~\cite{Abushammala2023}. 

\section{Phase diagram of $\text{Ba}\text{Co}\text{S}_2$}

\begin{figure}[tbh!]
\centering
 \includegraphics[width=\linewidth]{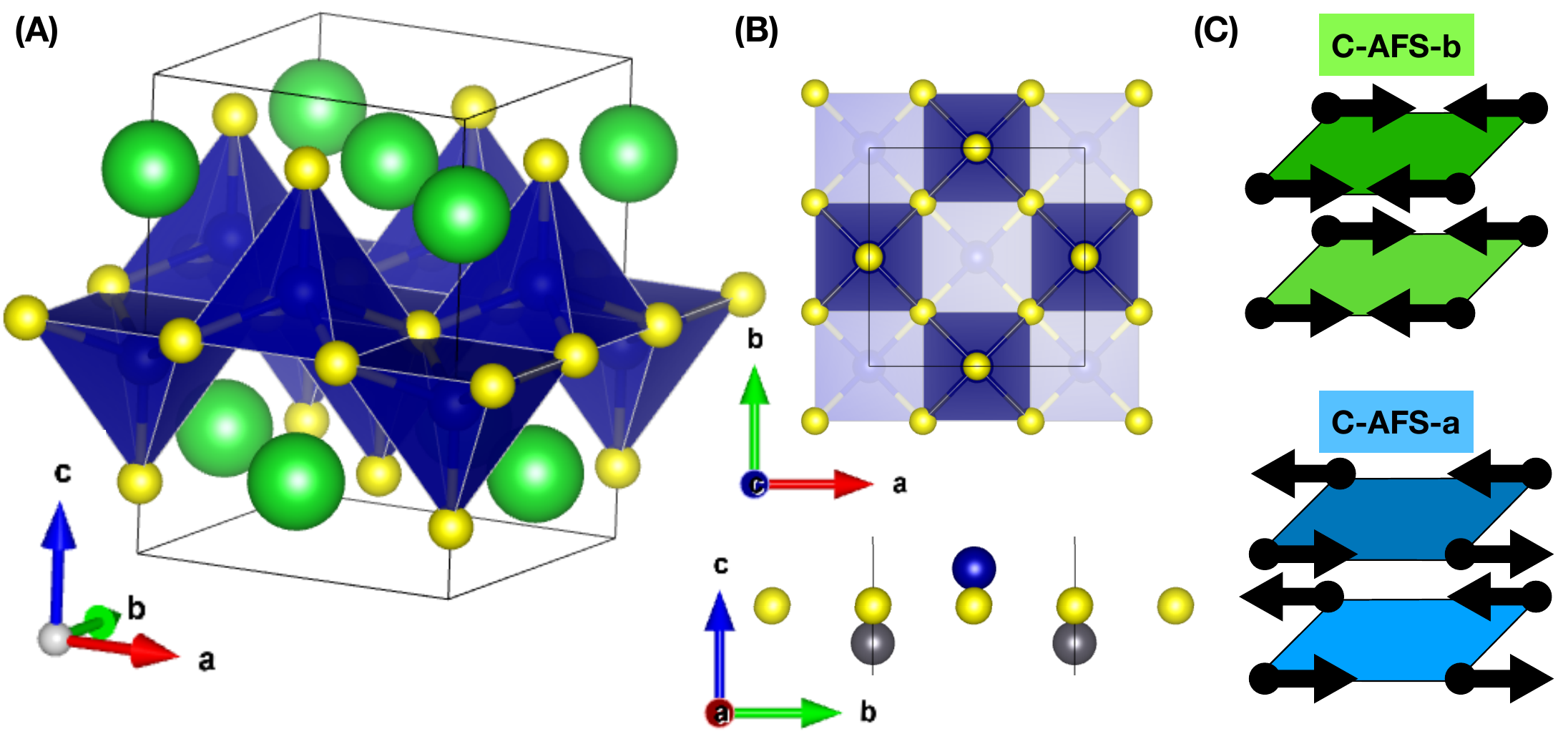} 
 \caption{(Color online) 
(A) Three-dimensional view of  BaCoS$_2$ tetragonal crystal structure. Ba atoms are the large green spheres, while S atoms are in yellow. The cobalt atoms sit inside the blue square based pyramids. (B) Top and lateral view of the structure, respectively, top and bottom panels. Note that, since the apexes of nearest neighbours pyramids point 
in opposite directions, there are two inequivalent Co atoms, shown as blue and grey spheres in the bottom panel, 
with opposite vertical displacements from the $a-b$ plane, which are connected by the non-symmorphic symmetry.
(C) Magnetic order in the low-temperature orthorhombic phase. Within each $a-b$ plane the spins form 
a striped antiferromagnet (AFS) with ferromagnetic chains coupled antiferromagnetically. The ferromagnetic chains can be either along $a$ (AFS-a) or along $b$ (AFS-b). The planes are stacked ferromagnetically, C-type stacking, thus 
the two equivalent configurations C-AFS-a and C-AFS-b. 
}
 \label{fig:Mag}
\end{figure}

$\text{Ba}\text{Co}\text{S}_2$ is a metastable layered compound that, 
quenched from high temperature, crystallises in an orthorhombic structure
with space group  $Cmme$, no. 67~\cite{Snyder1994}, characterised by in-plane primitive lattice vectors $\ba\not=\bb$. However, we believe physically more significant to consider as reference structure the higher-symmetry non-symmorphic $P4/nmm$ tetragonal one, thus $\ba=\bb$, of 
the opposite end member, BaNiS$_2$, and regard the orthorhombic distortion, 
which may occur through either $b>a$ or vice versa, as 
an instability driven by the substitution of Ni with the more correlated Co. 
The hypothetical tetragonal phase of $\text{Ba}\text{Co}\text{S}_2$ is 
shown in Fig.~\ref{fig:Mag}(A). Each CoS $a-b$ plane has two inequivalent cobalt atoms, Co(1) and Co(2), see Fig.~\ref{fig:Mag}(B), which are related to each other by the non-symmorphic symmetry. 

Below $T_N\simeq 290~\text{K}$~\cite{Abushammala2023}, a striped antiferromagnetic (AFS) phase sets in. 
In the $a-b$ plane it consists of ferromagnetic chains, either along $a$ (AFS-a) or $b$ (AFS-b), coupled 
antiferromagnetically, see Fig.~\ref{fig:Mag}(C). The stacking between the planes is C-type, i.e., ferromagnetic, 
thus the labels C-AFS-a and C-AFS-b that we shall use, as well as 
G-AFS-a and G-AFS-b whenever we discuss the G-type configurations with 
antiferromagnetic stacking.
We mention that the orthorhombic distortion with $b>a$ is associated with C-AFS-a, i.e., ferromagnetic bonds along $a$, 
and vice versa for $b<a$~\cite{Mandrus1997}, at odds with the expectation that ferromagnetic bonds are longer 
than antiferromagnetic ones. This counterintuitive behaviour represents a key test for the \textit{ab initio} 
calculations that we later present. 

Let us now briefly discuss how the electronic properties change across the magnetic transition. 
Above $T_N$, BaCoS$_2$ is a very bad conductor with no evidence of a Drude peak and with an optical conductivity that grows linearly in frequency up to around 1~eV~\cite{SantosCottin2018}. The dc resistivity shows an activated behaviour 
with activation energy $\simeq 91~\text{meV}$ that persists above $T_N$ up to 
$400~\text{K}$~\cite{SantosCottin2018}. 
At the N\'eel transition, a reduction of low-energy spectral weight starts, which is compensated by an increase 
at an energy of about $40~\text{meV}$~\cite{SantosCottin2018}. Such $40~\text{meV}$-gap opening occurs gradually below $T_N$.  On the contrary, 
the supposed Mott or charge-transfer peak at $\sim 1~\text{eV}$~\cite{Takenaka-PRB2001,SantosCottin2018}
is rather insensitive to the transition. These evidences suggest that, irrespective of the system being 
a poor metal or a weak semiconductor above $T_N$, there is an abundance of low energy excitations both 
above the N\'eel transition as well as below in the insulating phase.  

\begin{figure}
\centering
\includegraphics[width=0.45\textwidth]{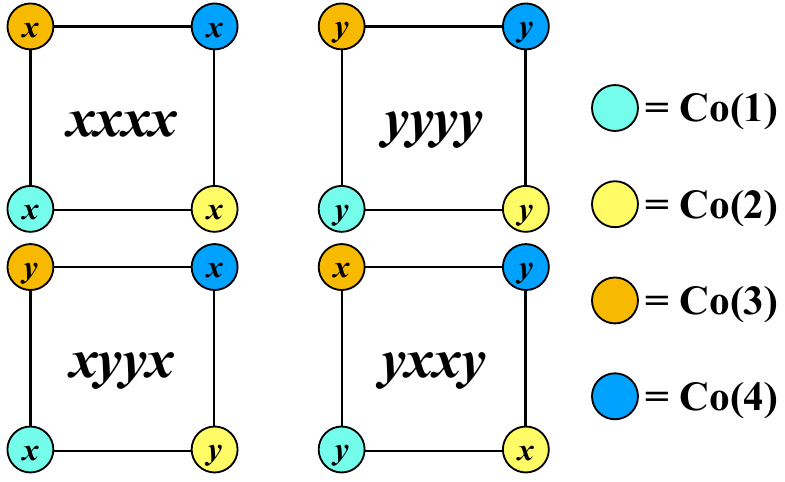}
\caption{Orbital arrangements allowed by the $Cmme$ space group. Co(1), Co(2), Co(3) and Co(4) correspond to 
the $4g$ Wyckoff positions occupied by the cobalt atoms. The label $x$($y$) indicates that the hole occupies 
the $d_{xz}$($d_{yz}$) orbital of the 3/4-filled $d_{xz}/d_{yz}$ doublet of the corresponding cobalt atom.}
\label{orbital-orders}
\end{figure}

Neutron scattering refinement and magnetic structure modelling in the low-temperature phase point to an ordered moment of $\mu_{Co}\sim 2.63-2.9 \mu_B$ \cite{Mandrus1997,Kodama1996}, suggesting that each Co$^{2+}$ is in a  
$S=3/2$ spin configuration, in agreement with the high-temperature magnetic susceptibility~\cite{Mandrus1997}.
Moreover, the form factor analysis of the neutron diffraction data~\cite{Kodama1996} indicates that the 
three 1/2-spins lie one in the $d_{3z^2-r^2}$, the other in the $d_{x^2-y^2}$, and the third either in the 
$d_{xz}$ or $d_{yz}$ $3d$-orbitals of Co. Since $d_{xz}$ and $d_{yz}$, which we hereafter denote shortly 
as $x$ and $y$ orbitals, form in the $P4/nmm$ tetragonal structure a degenerate $E_g$ doublet occupied by a single hole, such degeneracy is going to be resolved at low-temperature.  
That hints at the existence of some kind of orbital order, besides the spin one, in the magnetic orthorhombic phase. 
Let us try to anticipate by symmetry arguments which kind of order can be stabilised. \\
We observe that in the $Cmme$ orthorhombic structure the cobalt atoms occupy the Wyckoff positions $4g$, which, for convenience, we denote 
as $\text{Co}(1)\equiv (0,0,z)$, $\text{Co}(2)\equiv (1/2,0,-z)$, $\text{Co}(3)\equiv (0,1/2,-z)$, $\text{Co}(4)\equiv (1/2,1/2,z)$, and have symmetry $mm2$. As a consequence, the hole must occupy either the $x$ orbital or the $y$ one, but not a linear combination, and the chosen orbital must be the same for Co(1) and Co(4), as well as for Co(2) and Co(3). 
Therefore, if we denote as $d_n$, $d=x,y$, the orbital occupied by the hole on Co($n$), $n=1,\dots,4$, and as 
$d_1\,d_2\,d_3\,d_4$ a generic orbital configuration, then there are only four of them that are symmetry-allowed: 
$xxxx$, $yyyy$, $xyyx$ and $yxxy$, see Fig.~\ref{orbital-orders}. 
We remark that the density matrix of a unit cell in the lattice model can have finite diagonal components for all four configurations, while off-diagonal terms are prohibited by symmetry. 
We further note that $xxxx$ is degenerate with $yyyy$ in the tetragonal phase.
The choice of either of them is associated with the same $C_4\to C_2$ symmetry breaking 
that characterises both the AFS-a or AFS-b spin order and the orthorhombic distortion, $b>a$ or $a>b$. 
All these three choices can be associated with three Ising variables $\tau$, $\sigma$ and $X$ such that 
$\tau=+1$ corresponds to $xxxx$, $\sigma=+1$ to AFS-a, $X=+1$ to $b>a$, and vice versa. Since they all have the same 
symmetry, odd under $C_4$, they would be coupled to each other should we describe the transition by a Landau-Ginzburg 
functional. We shall hereafter denote as $Z_2(C_4)$ the Ising sector that describes the $C_4\to C_2$ symmetry breaking. \\
The other two allowed orbital configurations $xyyx$ and $yxxy$, see  Fig.~\ref{orbital-orders}, are instead degenerate both in the tetragonal and orthorhombic phases, but break 
the non-symmorphic symmetry (NS) that connects, e.g., Co(1) with Co(2) and Co(3). We can therefore associate 
to those configurations a new Ising sector $Z_2(\text{NS})$. 

We emphasise that the above conclusions rely on the assumption of a 
$Cmme$ space group. A mixing between $x$ and $y$ orbitals is instead 
allowed by the $Pba2$ space group proposed in Ref.~\cite{Schueller2020} 
as an alternative scenario for BaCoS$_2$ at room temperature. As a matter of fact, the two symmetry-lowering routes, $P4/nmm\to Cmme$ and $P4/nmm\to Pba2$, correspond to different Jahn-Teller-like 
distortions involving the $d_{xz}$-$d_{yz}$ doublet and the $E_g$ phonon mode
of the $P4/nmm$ structure at the $\mathbf{M}$ point, which is found to have imaginary frequency by \textit{ab initio} calculations~\cite{Schueller2020}.   
However, the latest high-accuracy X-ray diffraction data of Ref.~\cite{Abushammala2023} confirm the $Cmme$ orthorhombic structure even at room temperature, thus supporting our assumption. 

\section{\textit{Ab initio} analysis} 
We carried out \textit{ab initio} DFT and DFT+U calculations using the Quantum ESPRESSO package \cite{QE1,QE2}. 
The density functional is of GGA type, namely the Perdew-Burke-Ernzerhof (PBE) functional \cite{PBE}, on which local Hubbard interactions were added in case of the GGA+U fully rotational invariant framework \cite{GGAU1,GGAU2}.
If not stated otherwise, the geometry of the unit cell and the internal coordinates of the atomic positions in the orthorhombic structure were those determined experimentally, taken from Ref.~\onlinecite{Snyder1994}. 
For paramagnetic calculations, the relative atomic positions were kept fixed and the in-plane lattice constants $a=b$ chosen such that the unit cell volume matched the one of the orthorhombic structure.
Co and S atoms are described by norm-conserving pseudopotentials (PP) with non-linear core corrections, Ba atoms are described by ultrasoft pseudopotentials. 
The Co PP contains 13 valence electrons (3s$^2$,3p$^6$,3d$^7$), the Ba PP 10 electrons (5s$^2$,5p$^6$,6s$^2$), and S PPs are in a (3s$^2$,3p$^3$) configuration.
The plane-waves cutoff has been set to $120$Ry and we used a Gaussian smearing of $0.01$Ry.
The $k$-point sampling of the electron-momentum grid was at least $8\times8\times8$ points.\\
To determine the band structure and derive an effective low-energy model, we performed a Wannier interpolation with maximally localized Wannier functions (MLWF) \cite{MLWF1,MLWF2} using the 
Wannier90 package \cite{Wannier90}.\\

\noindent
In the first place, we checked if the tetragonal phase is unstable towards magnetism, considering both a 
conventional N\'eel order (AFM) compatible with the bipartite lattice and the observed AFS. We found, using $U=2.8~\text{eV}$ as motivated by a cRPA analysis \cite{David2018}, that the lowest energy state is indeed the AFS, the AFM and paramagnetic phases lying above by about 0.5~eV and 2.3~eV, respectively.\\
Let us therefore restrict our analysis to AFS and stick to $U=2.8~\text{eV}$. We use an 8-site unit cell that includes two planes, which allows us to compare C-AFS with G-AFS. 
In addition, we consider both the tetragonal structure with AFS-a, since AFS-b is degenerate, and the orthorhombic structure with $b>a$, 
in which case we analyse both AFS-a and AFS-b. For all cases, we consider all 
four symmetry-allowed orbital configurations, $xxxx$, $yyyy$, $xyyx$ and $yxxy$, 
assuming either a C-type or G-type orbital stacking between the two planes of the unit cell, so that, for instance, $\text{G}(xxxx)$ means that one plane is in the $xxxx$ configuration and the other in the $yyyy$ one. 
\begin{table}
\begin{tabular}{|c|c|c|}\hline
~spin and orbital configurations~ & ~$E$(Kelvin)~ & \# \\ \hline
C-AFS-a-C($xyyx$) & 0 & T0\\ \hline
G-AFS-a-C($xyyx$) & 2 & T1\\ \hline
C-AFS-a-G($xyyx$) & 14 & T2\\ \hline
G-AFS-a-G($xyyx$) & 22 & T3\\ \hline
G-AFS-a-G($xyxy$) & 50 & T4\\ \hline
C-AFS-a-G($xyxy$) & 52 & T5\\ \hline
C-AFS-a-C($xyxy$) & 52 & T6\\ \hline
C-AFS-a-C($yyyy$) & 57 & T7\\ \hline
G-AFS-a-C($xyxy$) & 64 & T8\\ \hline
G-AFS-a-C($yyyy$) & 73 & T9\\ \hline
C-AFS-a-G($xxyy$) & 79 & T10\\ \hline
C-AFS-a-G($yyyy$) & 86 & T11\\ \hline
G-AFS-a-G($yyyy$) & 89 & T12\\ \hline
G-AFS-a-C($xxyy$) & 89 & T13\\ \hline
C-AFS-a-C($xxyy$) & 93 & T14\\ \hline
G-AFS-a-G($xxyy$) & 95 & T15\\ \hline
G-AFS-a-C($xxxx$) & 171 & T16\\ \hline
C-AFS-a-C($xxxx$) & 176 & T17\\ \hline
\end{tabular}
\caption{DFT+U, with $U=2.8~\text{eV}$, energies in Kelvin and per formula unit of the low-lying spin and orbital configurations in the tetragonal structure with an 8-site unit cell, assuming a magnetic order AFS-$a$, being degenerate with AFS-$b$. The lowest energy state sets the zero of energy. Note that some 
states are doubly degenerate, for instance $C(xyyx)$ is degenerate with $C(yxxy)$
as well as $G(yyyy)$ is degenerate with $G(xxxx)$, and thus we just indicate one of them. 
Moreover, the table includes also configurations not allowed by the $Cmme$ orthorhombic space group, which, nonetheless, represent alternative symmetry-breaking paths from the tetragonal structure. Each state is labelled by T$n$, T referring to the tetragonal phase and $n$ being the ascending order in energy.}  
\label{Energies-tetragonal}
\end{table}
In Table~\ref{Energies-tetragonal} we report the energies per formula unit of several possible configurations in the tetragonal structure, including those that would be forbidden in the orthorhombic one. All energies are measured with respect to the lowest energy state and are expressed in Kelvin. In agreement with experiments, the lowest energy state T0 has spin order 
C-AFS, a or b being degenerate. In addition, it has C-type antiferro-orbital order, 
C$(xyyx)$. 
Remarkably, its G-type spin counterpart T1 is only 2~K above, supporting our observation that 
$T_N=290~\text{K}$ is anomalously large if compared to these magnetic excitations. 
\\
The energy differences between C-type orbital stacked configurations and their G-type counterparts are too small and variable to allow obtaining a reliable modelling of the inter-plane orbital coupling. \\
On the contrary, the energy differences between in-plane orbital configurations 
can be accurately reproduced by a rather simple modelling. We assume on each Co-site 
an Ising variable $\tau_3$ equal to the difference between the hole occupations of orbital $x$ and of orbital $y$. The Ising variable on a given site is coupled 
only to those of the four nearest neighbour sites in the $a-b$ plane, 
with exchange constants $\Gamma_{1a} = \Gamma_1 + \sigma\,\delta\Gamma_1$ 
and  $\Gamma_{1b} = \Gamma_1 - \sigma\,\delta\Gamma_1$ along $a$ and $b$, respectively. In addition, the Ising variables feel a uniform field $B_\tau\,\sigma$. Here, $\sigma$ is the Ising $Z_2(C_4)$ order parameter that distinguishes AFS-a, $\sigma=+1$ 
from AFS-b, $\sigma=-1$. 
\begin{table}
\begin{tabular}{|c|c|c|}\hline
orbital configuration & $E$ & $\Delta E$ \\ \hline
$xyyx$ & $-2\Gamma_1$ & 0\\ \hline
$xyxy$ & $-2\sigma\delta\Gamma_1$ & $2\Gamma_1-2\sigma\delta\Gamma_1$\\ \hline
$xxyy$ & $2\sigma\delta\Gamma_1$ & $2\Gamma_1+2\sigma\delta\Gamma_1$\\ \hline
$xxxx$ & $2\Gamma_1-\sigma\,B_\tau$ & $4\Gamma_1-\sigma\,B_\tau$ \\ \hline
$yyyy$ & $2\Gamma_1+\sigma\,B_\tau$ & $4\Gamma_1+\sigma\,B_\tau$ \\ \hline
\end{tabular}

\vspace{0.2cm}

\begin{tabular}{|c|c|c|c|}\hline
~ & $\Gamma_1(\text{K})$ & $\delta\Gamma_1(\text{K})$ & $B_\tau(\text{K})$ \\ \hline
C-AFS & $33\pm 4$ & 10 & 60 \\ \hline
G-AFS & $31 \pm 4$ & 6 & 49 \\ \hline
\end{tabular}
\caption{Upper table: the energies of the different orbital configurations 
within the assumed nearest neighbour antiferromagnetic Ising model 
with exchange constants $\Gamma_1+\sigma\,\delta\Gamma_1$ along $a$, 
$\Gamma_1-\sigma\,\delta\Gamma_1$ along $b$, and uniform pseudo-magnetic field 
$\sigma\,B_\tau$. Bottom table: the values of those parameters extracted 
through Table~\ref{Energies-tetragonal}. We just consider the C-type orbital 
stacked configurations, since the G-type ones do not allow fixing $B_\tau$. }
\label{parameters-tetragonal}
\end{table}
We find that the spectrum is well reproduced 
by the parameters in Table~\ref{parameters-tetragonal}.\\

We now move to the physical orthorhombic structure, assuming $b>a$ with $b/a = 1.008$, and recalculate 
all above energies but considering only the orbital configurations 
allowed by the $Cmme$ space group. In this case, we have to distinguish 
between AFS-a and AFS-b, which are not anymore degenerate. The results are shown 
in Table~\ref{Energies-orthorhombic}.
\begin{table}
\begin{tabular}{|c|c|c|}\hline
~spin and orbital configurations~ & ~$E$(Kelvin)~ & \# \\ \hline
C-AFS-a-C($xyyx$) & 0  & O0 \\ \hline
G-AFS-a-C($xyyx$) & 2 & O1 \\ \hline
C-AFS-a-G($xyyx$) & 14 & O2 \\ \hline
C-AFS-b-C($xyyx$) & 20 & O3 \\ \hline
G-AFS-a-G($xyyx$) & 22 & O4 \\ \hline
G-AFS-b-C($xyyx$) & 22 & O5 \\ \hline
C-AFS-b-G($xyyx$) & 34 & O6 \\ \hline
G-AFS-b-G($xyyx$) & 42 & O7 \\ \hline
C-AFS-b-C($xxxx$) & 50 & O8 \\ \hline
G-AFS-b-C($xxxx$) & 65 & O9 \\ \hline
C-AFS-b-G($xxxx$) & 79 & O10 \\ \hline
G-AFS-b-G($xxxx$) & 82 & O11 \\ \hline
C-AFS-a-C($yyyy$) & 85 & O12 \\ \hline
C-AFS-a-G($xxxx$) & 93 & O13 \\ \hline
G-AFS-a-G($xxxx$) & 96 & O14 \\ \hline
G-AFS-a-C($yyyy$) & 101 & O15 \\ \hline
G-AFS-a-C($xxxx$) & 160 & O16 \\ \hline
C-AFS-a-C($xxxx$) & 165 & O17 \\ \hline
G-AFS-b-C($yyyy$) & 203 & O18 \\ \hline
C-AFS-b-C($yyyy$) & 209 & O19 \\ \hline
\end{tabular}
\caption{Same as in Table~\ref{Energies-tetragonal} but for the orthorhombic structure with $b>a$, $b/a=1.008$. In this case, AFS-a and AFS-b are not degenerate, and thus both have been studied. Only the orbital configurations allowed by symmetry 
are shown. The states are labelled by O$n$, where O refers to the orthorhombic phase and $n$ is the order.
}
\label{Energies-orthorhombic}
\end{table}
We remark that the \textit{ab initio} calculation correctly predicts that the lowest energy state O0 has ferromagnetic bonds along $a$ despite $b>a$, which, as we mentioned, is an important test for the theory. 
The energy difference between AFS-a and AFS-b, e.g., O0 and O3, is about 20~K, 
and gives a measure of the spin-exchange spatial anisotropy in the $a$ and $b$ directions due to the orthorhombic distortion. This small value implies 
that the N\'eel transition temperature $T_N\simeq 290~\text{K}$ is 
largely insensitive to the orthorhombic distortion that exists also 
above $T_N$~\cite{Mandrus1997,Abushammala2023}. 
In particular, the energy difference between C-type and G-type stacking, 
O0 and O1 in Table~\ref{Energies-orthorhombic}, remains the same tiny value 
found in the tetragonal phase. Table~\ref{Energies-orthorhombic} thus suggests that the spin configurations C-AFS-a, C-AFS-b, G-AFS-a and 
G-AFS-b are almost equally probable at the N\'eel transition, and that despite 
the orthorhombic structure. \\
Concerning the effective Ising model that controls 
the in-plane orbital configuration, we find that the parameter that is most affected by the orthorhombic distortion is $B_\tau$, which decreases in the 
AFS-a configurations and increases in AFS-b ones. \\
The calculated magnetic moment per Co atom in the lowest energy state, 
O0 in Table~\ref{Energies-orthorhombic}, is $\mu_{AFS}\sim2.65~\mu_B$, 
in quite good agreement with experiments~\cite{Mandrus1997,Kodama1996}. 
We mention that the actual magnetic moment of Co is lower, $\mu_{AFS}^{Co}\sim2.29~\mu_B$. 
This difference is due to the strong Co-S hybridization that yields a non-negligible magnetic moment on the sulfur sites.

\subsection{Orthorhombic distortion.} 
\begin{figure}[tb!]
\centering
 \includegraphics[width=\linewidth]{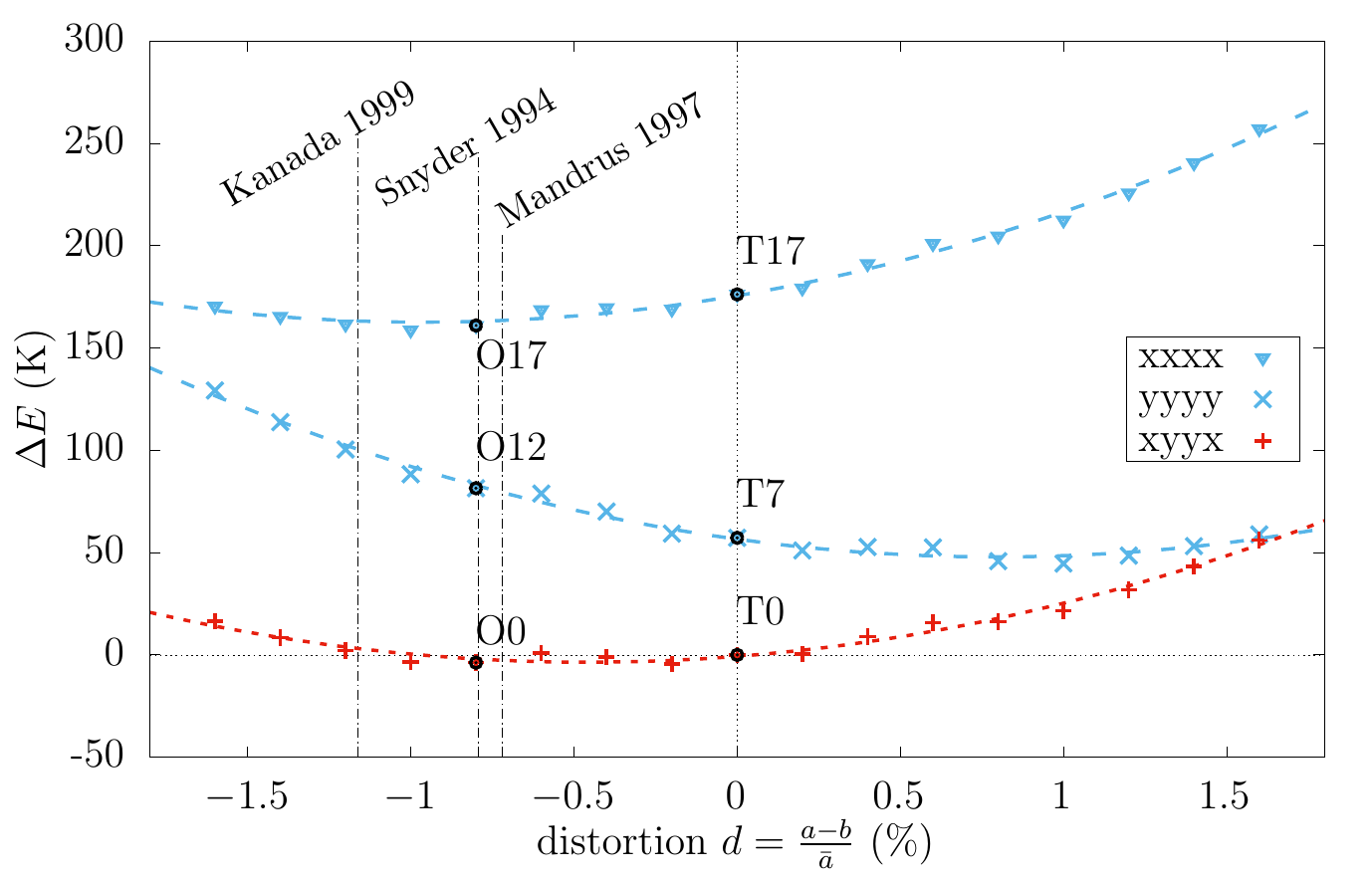} 
 \caption{(Color online) 
	Energy $\Delta E$ per formula unit as function of the orthorhombic distortion $d$ at fixed unit cell volume and measured with respect to the C($xyyx$) phase at $d=0$ (T0 in Table~\ref{Energies-tetragonal}). We compare $C(xyyx)$ with the two nematic configurations 
	$C(xxxx)$ and $C(yyyy)$ assuming a magnetic order C-AFS-a. 
	Experimental data are taken from Refs.~\onlinecite{Snyder1994,Mandrus1997,Kanada1999}.
	The phases T0, T7, and T17 refer to Table~\ref{Energies-tetragonal}, phases O0, O12, and O17 to Table~\ref{Energies-orthorhombic}. 
}
 \label{fig:dist}
\end{figure}

BaCoS$_2$ has a small orthorhombic distortion quantified by $d=\frac{a-b}{\bar{a}}\sim 0.8\%$ \cite{Snyder1994} with $\bar{a}=\frac{a+b}{2}$, which is further reduced to $d\sim0.4\%$ under high pressure synthesis \cite{Abushammala2023}.
Therefore, the stability of the different phases with respect to such an orthorhombic distortion can be regarded as a further criterion to identify the correct electronic ground state of the system.
Whereas the PM phase favours a strong orthorhombic deformation of $d\sim8\%$, the magnetic stripe phases have minimal energy variations for small distortions. 
The precise values differ between the different orbital configurations as shown in Fig.~\ref{fig:dist}, and the lowest energy is found for the orbital ordered configuration $xyyx$ at $d^{xyyx}_{\mathrm{min}}\sim0.5\%$.
For the nematic $xxxx$ configuration, the distortion is in good agreement with the experimental values  ($d^{xxxx}_{\mathrm{min}}\sim-1.0\%$) albeit being much higher in energy.
The energy minimum of the nematic $yyyy$ phase is competing with $xyyx$, but located at positive instead of negative distortion ($d^{yyyy}_{\mathrm{min}}\sim0.8\%$).
\\ 

\subsection{Insulator-to-metal transition.} 
\begin{figure}[tb!]
\centering
 \includegraphics[width=\linewidth]{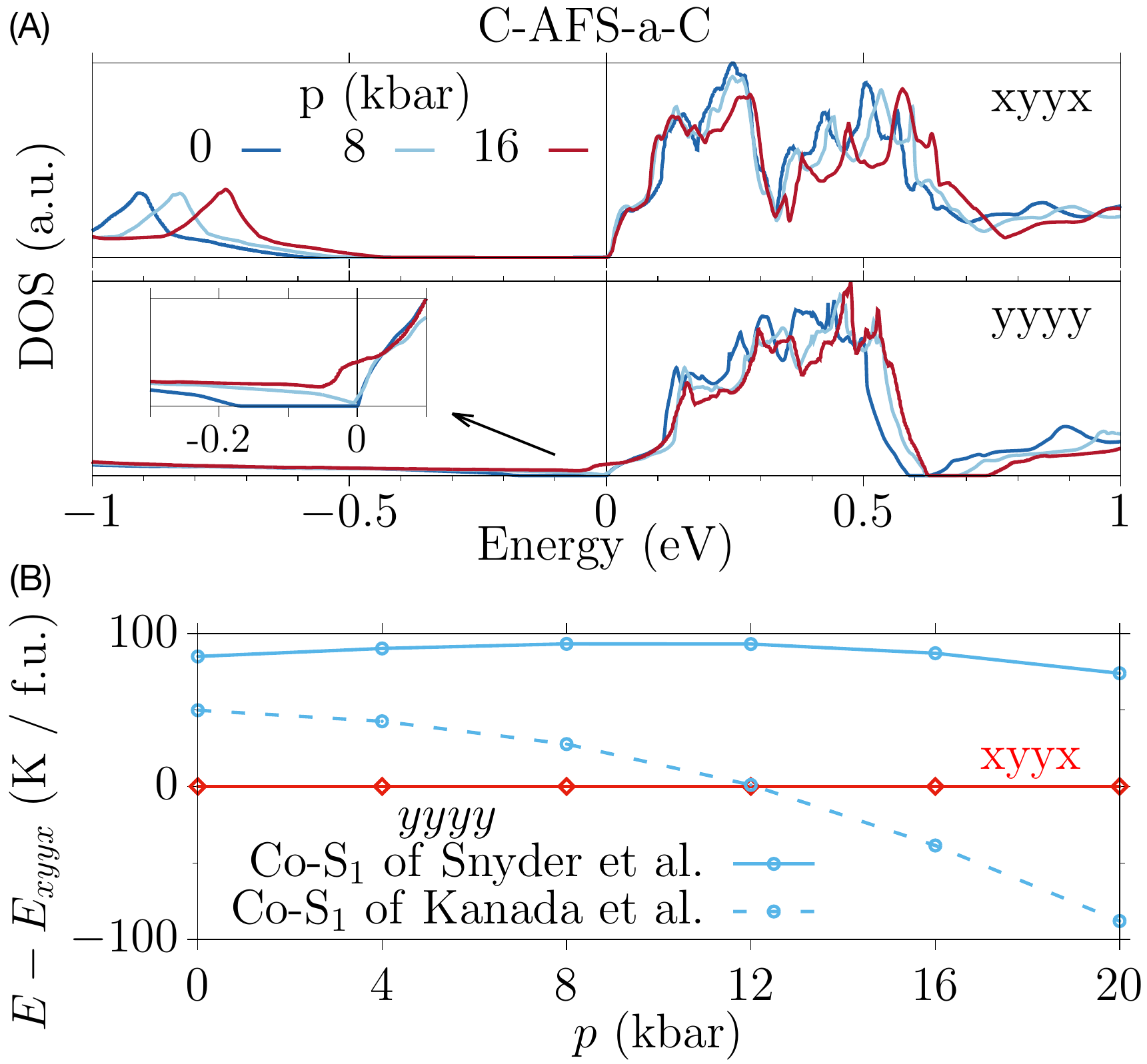} 
 \caption{(Color online) 
 	(A) Density of states in the competing C-AFS-a-C phases for the structure taken from Ref.~\onlinecite{Snyder1994} at different pressure with orbital nematic (\textit{yyyy}) and anti-ferro orbital order (\textit{xyyx}).
	The energy gap of the \textit{yyyy} phase is closing at $p\gtrsim8$ kbar.
    (B) Relative energy of these two phases within the structure taken from Snyder \textit{et al.}~\cite{Snyder1994} as well as with the reduced apical sulfur distance Co-S$_1$ of Kanada \textit{et al.}~\cite{Kanada1999}. 
}
 \label{fig:lowpress}
\end{figure}

The insulator-metal transition can be controlled either by chemical doping, for instance with Ni atoms \cite{Martinson1993}, or by applying pressure \cite{Kodama1996} .
We focus here on the latter case, where a critical pressure of $p_{cr}\sim 1.3$ GPa \cite{Sato1998,Yasui1999,Kanada1999,Guguchia2019} was found in experiments to be sufficient to render the system metallic.
Close to $p_{cr}$ the transition from AFM insulator to PM metal occurs via the formation of an intermediate antiferromagnetic metal phase. \\
Based on the structural changes under pressure reported in Ref.~\onlinecite{Kanada1999}, we performed DFT+U calculations for the low-energy solutions identified earlier~\footnote{In order to be consistent with the rest of the paper, we used the relative structural changes of the lattice constants under pressure reported in Ref.~\onlinecite{Kanada1999} and applied them to the ambient pressure data of Ref.~\onlinecite{Snyder1994}.}.
In Ref.~\onlinecite{Kanada1999} the structural parameters show a rather continuous evolution upon pressure within the metallic and insulating regimes. 
At the phase transition, however, the structural changes are large.
Nevertheless, one can parametrize the distinct low- and high-pressure regimes separately whose structural parameters vary as a function of pressure.
Here, we focus on the low-pressure parametrization and compare the energy gap of the two C-AFS-a competing solutions C(\textit{xyyx}) and C(\textit{yyyy}) as a function of pressure. In Fig.~\ref{fig:lowpress} we plot the density of states
(DOS) of both configurations for several pressures. We note that, while 
C(\textit{xyyx}) remains insulating at all pressures, the gap of the 
C(\textit{yyyy}) configuration closes around $p\simeq 8~\text{kbar}$. 
Such behaviour is robust either upon choosing the atomic positions of 
Ref.~\onlinecite{Kanada1999} or those of Ref.~\onlinecite{Snyder1994}. 
On the contrary, the relative energies of C(\textit{xyyx}) and C(\textit{yyyy}), 
which are extremely sensitive to the distance between Co and apical S, 
do depend on that choice.
Indeed, Ref.~\onlinecite{Kanada1999} and Ref.~\onlinecite{Snyder1994} report significantly different 
Co-apical S distances. 
The insulating C(\textit{xyyx}) remains the lowest energy state at all 
investigated pressures using the atomic positions of Ref.~\onlinecite{Snyder1994}, see top panel of Fig.~\ref{fig:lowpress}(B).   
Conversely, the reduced Co-apical S distance of the crystal structure reported in Ref.~\onlinecite{Kanada1999} predicts 
a first order transition at $p\sim12$~kbar 
between an insulating C(\textit{xyyx}) and a metallic C(\textit{yyyy}), 
consistent with the metal-insulator transition reported experimentally, see bottom panel. 
Should the latter scenario be representative of BaCoS$_2$ under pressure, 
it would imply that the metal-insulator transition is also accompanied 
by the recovery of the non-symmorphic symmetry. 

\subsection{Wannierization}
In order to gain further insight into the electronic structure of \BCS\ in general and, in particular, of the bands with pronounced $d_{xz}$ and $d_{yz}$ character, we generate two theoretical model Hamiltonians with maximally-localized Wannier functions for Co-$d$-like and Co-$d_{xz/yz}$-like orbitals respectively.
To this aim, we construct Wannier fits using the Wannier90 tool \cite{Wannier90} based on the paramagnetic DFT+U calculation using a $4\times4\times4$ k-grid with a doubled in-plane unit cell comprising 4 Co atoms.
\begin{figure}[hbt]
\includegraphics[width=0.5\textwidth]{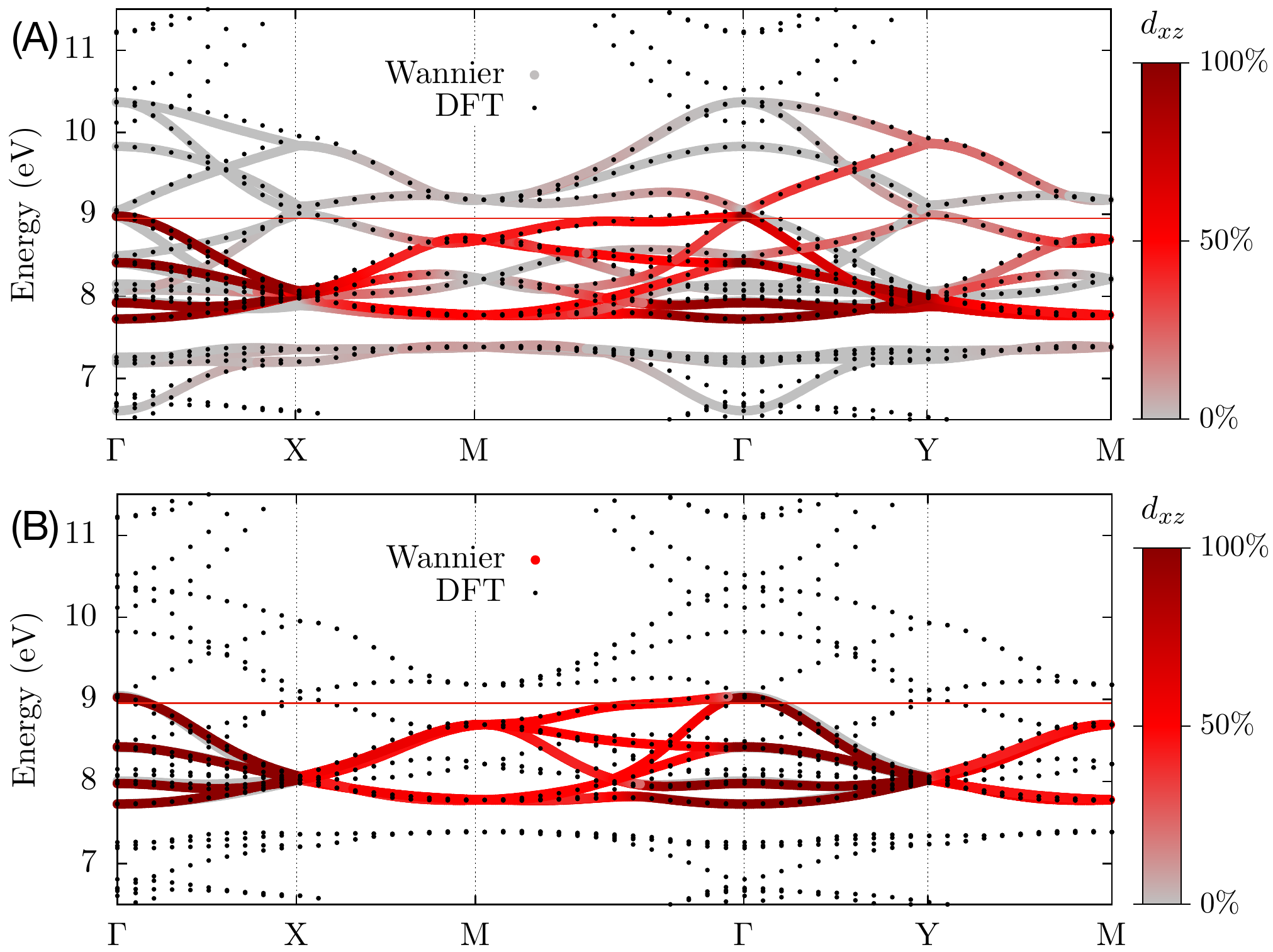}
\caption{DFT band structure and eigenstates of our Wannier Hamiltonians along the high-symmetry path $\Gamma-X-M-\Gamma-Y-M$: (A) Co-$d$ model and (B) $d_{xz}$-$d_{yz}$ model with projection onto the $d_{xz}$-orbital character in red.}
\label{fig:Wannier}
\end{figure}
For both models, the resulting Wannier Hamiltonian reproduces overall well the DFT band structure, see Fig.~\ref{fig:Wannier}.
Whereas the fit of the 5-orbital model is nearly perfect, the 2-orbital model shows small deviations along the $\mathbf{M}-\boldsymbol{\Gamma}$ direction due to missing hybridization with the other Co-$d$ orbitals.\\

\begin{table}[ht]
\begin{displaymath}
\begin{array}{|c|c|}\hline
\text{bond~direction} & \text{hopping~matrix~(meV)}\\ \hline
T_{(1,1,0)}=T_{(-1,-1,0)} & 
\begin{pmatrix}
96 & 102\\
102 & 94
\end{pmatrix}
\\ \hline
T_{(1,-1,0)}=T_{(-1,1,0)} & 
\begin{pmatrix}
96 & -102\\
-102 & 94
\end{pmatrix}
\\ \hline
T_{(1,0,0)}=T_{(-1,0,0)} & 
\begin{pmatrix}
2 & 0\\
0 & -43
\end{pmatrix}
\\ \hline
T_{(0,1,0)}=T_{(0,-1,0)} & 
\begin{pmatrix}
-48 & 0\\
0 & 2
\end{pmatrix}
\\ \hline
T_{(1,0,1)}=T_{(-1,0,1)} & 
\begin{pmatrix}
-68 & 0\\
0 & 18
\end{pmatrix}
\\ \hline
T_{(0,1,1)}=T_{(0,-1,1)} & 
\begin{pmatrix}
20 & 0\\
0 & -69
\end{pmatrix}\\ \hline
\end{array}
\end{displaymath}
\caption{
Leading hopping processes $T_{(n_x,n_y,n_z)}$, where $\br=(n_x,n_y,n_z)$ identifies the bond connecting Co(1), see Fig.~\ref{orbital-orders}, to another cobalt 
at distance $\br$. The bonds emanating from Co(2) are obtained by the 
non-symmorphic symmetry, which, in particular, implies $n_z\to -n_z$. 
All hopping processes are written as matrices in the subspace $\big(d_{xz},d_{yz}\big)$. The values, in meV, are obtained by the 5-orbital model 
restricted to the $\big(d_{xz},d_{yz}\big)$ subspace.}
\label{tab:Hops} 
\end{table}
Table~\ref{tab:Hops} shows the leading hopping processes of the 5-band model restricted to the $\big(d_{xz},d_{yz}\big)$ subspace. \\
We note that, because of the staggered shift of the Co atoms out of the sulfur basal plane, the largest intra-layer hopping is between next-nearest neighbour (NNN) cobalt atoms instead of nearest-neighbour (NN) ones. 
Moreover, the stacking of the sulfur pyramids and the position of the intercalated Ba atoms makes the inter-layer NN hopping negligible, contrary to the NNN one that 
is actually larger than the in-plane NN hopping, but still smaller than the in-plane NNN one. \\
We finally remark that the orthorhombic distortion has a very weak effect on the inter-layer hopping, which is consistent with the 
tiny energy difference between C-AFS and G-AFS being insensitive to the distortion, 
compare, e.g., the energies of T1 and O1 in Tables~\ref{Energies-tetragonal}
and \ref{Energies-orthorhombic}, respectively.

\section{Effective Heisenberg model}
\label{sec:Model}
\begin{figure}[hbt]
\centerline{\includegraphics[width=0.45\textwidth]{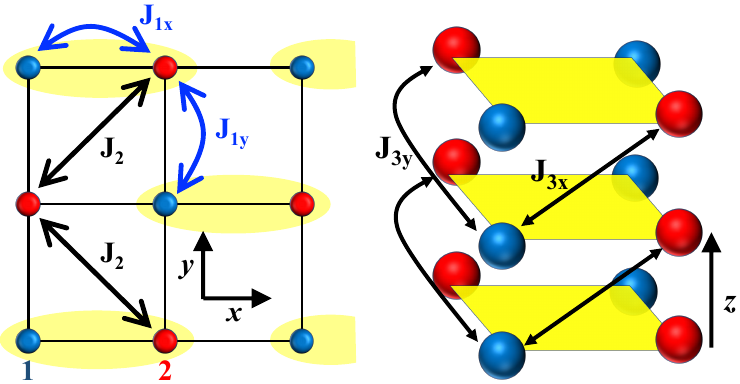}}
\vspace{-.5cm}
\centerline{\includegraphics[width=0.3\textwidth]{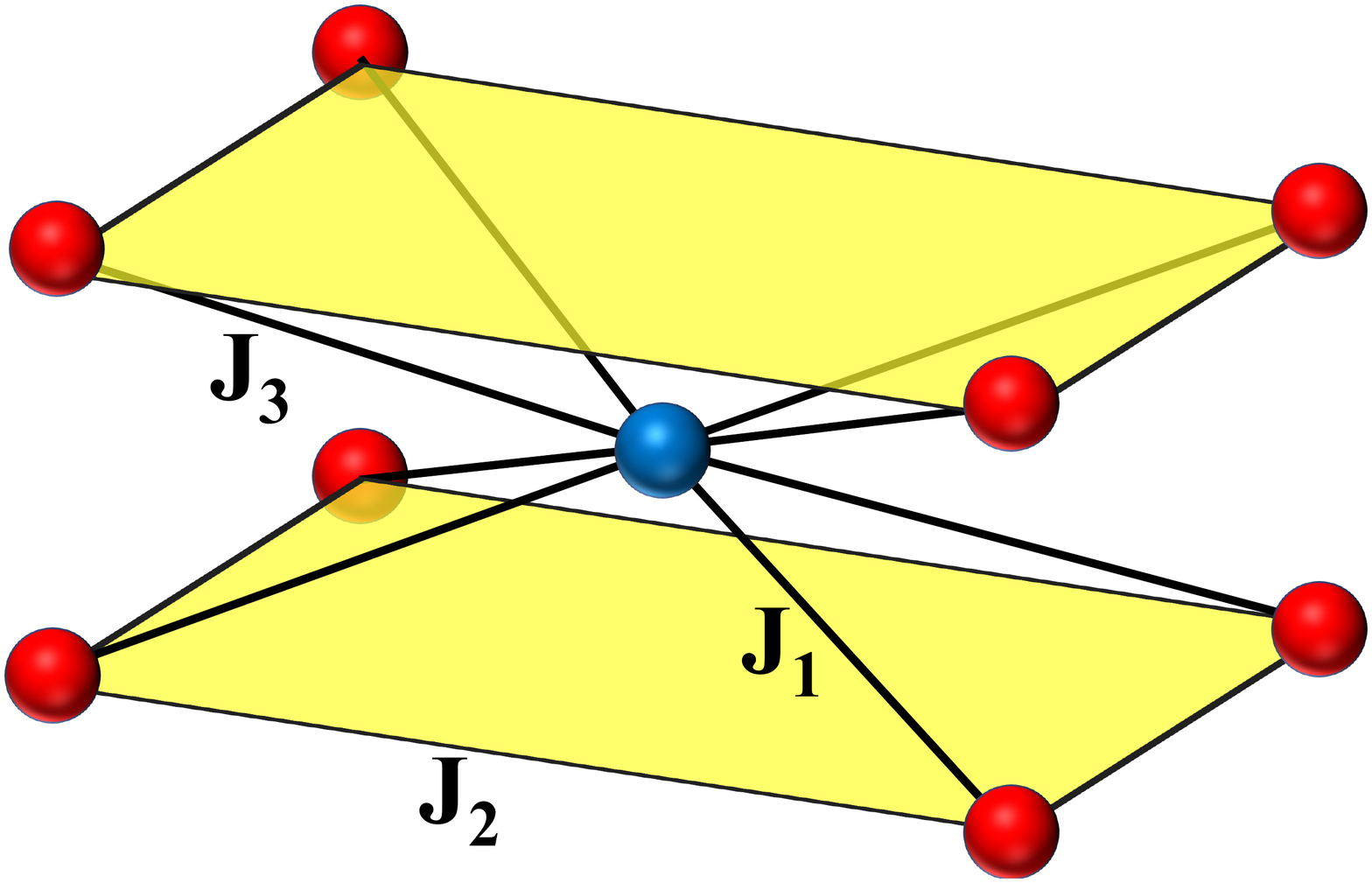}}
\vspace{-0.5cm}
\caption{Top panels: Heisenberg model for BaCoS$_2$. The left panel shows the 
$J_1-J_2$ model on the $a-b$ plane, while the right panel show how nearest neighbour planes are coupled to each other by the exchange $J_3$. 
The latter forces to deal with a two-site unit cell, highlighted in yellow in the 
left panel. This model is actually equivalent to the Heisenberg model on a bcc 
lattice shown in the bottom panel.} 
\label{J1-J2-model}
\end{figure}
We already mentioned that the largest energy scales 
that emerge from the \textit{ab initio} calculations 
are the magnetic ones separating the lowest-energy AFS configuration from 
the N\'eel and paramagnetic states.  
Therefore, even though BaCoS$_2$ seems not to lie deep inside a Mott insulating regime, we think it is worth discussing qualitatively the spin dynamics in terms 
of an effective $S=3/2$ Heisenberg model. If we assume that the leading contribution to the exchange constants derives from the hopping processes within the $d_{xz}-d_{yz}$ subspace, then Table~\ref{tab:Hops} suggests  
the Heisenberg model shown in Fig.~\ref{J1-J2-model}. 
According to this figure, the exchange constants $J_{1x/y},J_2,$ and $J_{3x/y}$ are related to the hopping terms $T_{(\pm1,0,0)/(0,\pm1,0)}, T_{\pm(1,-1,0)/\pm(1,1,0)},$ and $T_{(\pm1,0,1)/(0,\pm1,1)}$, reported in Table~\ref{tab:Hops}.
This model consists of frustrated $J_1-J_2$ planes~\cite{J1-J2-original,Moreo-PRB1990,Capriotti-PRL2004,Becca-PRB2005,Alberto-PRB2006} coupled to each other by a still frustrating $J_3$. 
In order to be consistent with the observed columnar magnetic order, the exchange constants have to satisfy the inequality $2J_2 > \left|J_3\right| +\left|J_1\right|$.
Moreover, $J_3$ forces to deal with a two sites unit cell, highlighted in yellow in the left panel of 
Fig.~\ref{J1-J2-model} where the non-equivalent cobalt sites are referred to as 
Co(1), in blue, and Co(2), in red, respectively. The reason is that Co(1) on a plane is only coupled to Co(2) on the plane above but not below, and vice versa for Co(2). \\
We write, for $a=1,3$,  
\beal
J_{ax}&=J_a\,\big(1-\delta_a\big)\,,&
J_{ay}&=J_a\,\big(1+\delta_a\big)\,,
\label{J-orthorhombic}
\eal
where, in analogy with the single-layer $J_1-J_2$ model~\cite{J1-J2-original}, 
$\delta_a\not=0$ are Ising order parameters associated with the $C_4\to C_2$ symmetry breaking. Moreover, we define the in-plane Fourier transforms  
\beal
\bd{S}_{\ell,n}(\bq) &= \sum_\bR\,\esp{-i\bq\cdot\bR}\;\bd{S}_{\ell,n,\bR}\,,
\eal
where $\bR$ labels the $N$ unit cells in the $a-b$ plane, 
$\ell=1,2$ the two sites 
(sublattices) within each unit cell, and $n=1,\dots,L$ the layer index. 
Similarly, we introduce the three-dimensional Fourier transforms
\bealn
\bd{S}_\ell(\bq,q_z) &= \sum_n\,\esp{-iq_z n}\; \bd{S}_{\ell,n}(\bq)\,,
\eal
as well as the spinor 
\bealn
\mathbf{S}(\bq,q_z) &= 
\begin{pmatrix}
\bd{S}_{1}(\bq,q_z)\\
\bd{S}_{2}(\bq,q_z)
\end{pmatrix}\,.
\eal
With those definitions, the Hamiltonian reads 
\bw
\beal
H &= \fract{1}{N}\,\sum_{n\bq}\bigg\{
J_2(\bq)\,\Big(\bS_{1,n}(\bq)\cdot\bS_{1,n}(-\bq)
+ \bS_{2,n}(\bq)\cdot\bS_{2,n}(-\bq)\Big)+J_1\, \Big(\gamma(\bq,\delta_1)\,\bS_{1,n}(\bq)\cdot\bS_{2,n}(-\bq) + H.c.\Big)\\
&\qquad\qquad\qquad  +J_3\, \Big(\gamma(\bq,\delta_3)\,\bS_{1,n}(\bq)\cdot\bS_{2,n+1}(-\bq)
+H.c.\Big)\bigg\} \equiv H_2 + H_{1} + H_3\\
&=\fract{1}{NL}\,\sum_{\bq,q_z}\,
\mathbf{S}^\dagger(\bq,q_z)\,
\begin{pmatrix}
J_2(\bq) & J_1\,\gamma(\bq,\delta_1)^* +J_3\,\gamma(\bq,\delta_3)^*\,\esp{-iq_z}\\
J_1\,\gamma(\bq,\delta_1) +J_3\,\gamma(\bq,\delta_3)\,\esp{iq_z} & 
J_2(\bq)
\end{pmatrix}\,\mathbf{S}(\bq,q_z)
\\
&\equiv \fract{1}{NL}\,\sum_{\bq,q_z}\,
\mathbf{S}^\dagger(\bq,q_z)\,\hat{J}(\bq,q_z)\,
\mathbf{S}(\bq,q_z)\,,\label{model: H_SS}
\eal
\ew
where $H_a$ is proportional to $J_a$, $a=1,2,3$, 
\bealn
J_2(\bq) &= 2J_2\,\cos q_x\,\cos q_y\,,\\
\gamma(\bq,\delta) &= \esp{iq_x}\,\Big[\big(1-\delta\big)\,\cos q_x + 
\big(1+\delta\big)\,\cos q_y\Big]\;.
\eal
The classical ground state corresponds to the modulation wavevector $\bQ=(\pi,0,Q_z)\equiv (0,\pi,Q_z)$, related to an antiferromagnetic order within each sublattice on each layer, yielding the lowest eigenvalue of $\hat{J}(\bq,q_z)$, i.e., 
\be
\lambda_{Q_z} = -2J_2 - 2\sqrt{ J_1^2 \delta_1^2 + 
J_3^2 \delta_3^2 + 2J_1 J_3 \delta_1 \delta_3 \cos Q_z\,}\,.
\label{lambda}
\ee
The expression of $\lambda_{Q_z}$ shows that inter-layer magnetic coherence sets in only when the two Ising-like order parameters, $\delta_1$ and $\delta_3$, 
lock together. Specifically, $J_1 J_3 \delta_1 \delta_3>0$ stabilises C-AFS, $Q_z=0$, otherwise G-AFS, $Q_z=\pi$. We already know that the former is lower in energy, though by only few Kelvins, see Table~\ref{Energies-orthorhombic}. Moreover, an orthorhombic distortion 
$b>a$ favours AFS-a, which implies $J_1\,\delta_1 + J_3\,\delta_3>0$, 
even though AFS-b is higher by only $20~\text{K}$ according to DFT+U, 
see O3 in Table~\ref{Energies-orthorhombic}.\\

\noindent
If we assume that the magnetic excitations in the AFS state of strained tetragonal BaCo$_{0.9}$Ni$_{0.1}$S$_{1.9}$ are representative of those of orthorhombic BaCoS$_2$, then the inelastic neutron scattering data 
of Ref.~\cite{Shamoto-PRR2021} at $200~\text{K}$ and for in-plane transfer momentum 
are consistent with $J_2\simeq 9.3~\text{meV}$, 
$J_1+J_3\simeq -2.34~\text{meV}$, and $J_1 \delta_1+J_3\delta_3\simeq 0.53~\text{meV}$. However, the scattering data for out-of-plane transfer momentum only yield an upper bound to the propagation velocity of the Goldstone mode along $z$, which actually corresponds to 
$4J_1J_3\delta_1 \delta_3< 0.08~\text{meV}^2$ for $J_3\simeq J_1$. 
Such small bound, around a quarter of $\left(J_1 \delta_1+J_3\delta_3\right)^2$, 
suggests that the two order parameters $\delta_1$ and $\delta_3$ are already formed at $80~\text{K}$ below $T_N\simeq 280~\text{K}$, whereas their mutual locking is still suffering from fluctuations. We finally observe that the ferromagnetic sign of $J_1$ and $J_3$ is consistent with the diagonal hopping matrices in the corresponding directions, see Table~\ref{tab:Hops}, and the antiferro-orbital order.\\

\noindent
To better understand the interplay between the $Z_2\left(C_4\right)$ Ising degrees of freedom and the magnetic order at $T_N$, we investigate in more detail 
the Hamiltonian \eqn{model: H_SS} with $\delta_1=\delta_3=0$, thus 
$J_{1x}=J_{1y}=J_1$ and $J_{3x}=J_{3y}=J_3$.
Since $J_2>0$ is the dominant exchange process, the classical ground state corresponds to 
the spin configuration 
\bealn
\bS_{i, n}(\bq) &= NS\,\bd{n}_{3,i, n}\,\delta_{\bq,\bQ}\,,& 
i &= 1,2\,,& n&=1,\dots,L\,,
\eal
where $S=3/2$ is the spin magnitude, $\bd{n}_{3,i,n}$ is a unit vector, and 
\beal
\bQ &=\big(\pi,0,0)\equiv (0,\pi,0)\,,
\eal
the equivalence holding since $\bd{G}=(\pi,\pi,0)$ is a 
primitive lattice vector for the two-site unit cell. 
In other words, each sublattice on each plane is N\'eel ordered, and its staggered 
magnetisation $\bd{n}_{3,i, n}$ is arbitrary. We therefore expect that quantum and thermal fluctuations 
may yield a standard order-from-disorder phenomenon~\cite{Villain-1980}.\\
Within spin-wave approximation, the spin operators can be written as 
\bealn
\bS_{i,n}(\bq)\cdot\bd{n}_{3, i,n} &\simeq NS\,\delta_{\bq,\bQ} - \Pi_{i,n}(\bq-\bQ)\,,\\
\bS_{i,n}(\bq)\cdot\bd{n}_{1, i,n} &\simeq \sqrt{NS\,}\; x_{i,n}(\bq)\,,\\  
\bS_{i,n}(\bq)\cdot\bd{n}_{2, i,n} &\simeq \sqrt{NS\,}\; p_{i,n}(\bq-\bQ)\,,
\eal
where $\bd{n}_{1, i,n}$, $\bd{n}_{2, i,n}$ and $\bd{n}_{3, i,n}$ 
are orthogonal unit vectors, $x_{i,n}^\dagger(\bq) = x^\dagga_{i,n}(-\bq)$ and $p_{i,n}^\dagger(\bq) = p^\dagga_{i,n}(-\bq)$ are conjugate 
variables, i.e., 
\bealn
\Big[\,x^\dagga_{i,n}(\bq)\,,\,p_{j,m}^\dagger(\bq')\,\Big] &= i\,\delta_{i,j}\,\delta_{n,m}\,\delta_{\bq,\bq'}\,,
\eal
and
\bealn
\Pi_{i,n}(\bq-\bQ) &= \fract{1}{2}\sum_\bk\Big(
x_{i,n}^\dagger(\bk)\,x^\dagga_{i,n}(\bk+\bq-\bQ) \\
&\qquad\qquad + p_{i,n}^\dagger(\bk)\,p^\dagga_{i,n}(\bk+\bq-\bQ) - \delta_{\bq,\bQ}\Big)\,.
\eal
The three terms of the Hamiltonian \eqn{model: H_SS} thus read, at leading order in quantum fluctuations, i.e., in the harmonic approximation,  
\bealn
H_2 &\simeq  E_0+ S\sum_{i,n\,\bq}\Big(J_2(\bq)-J_2(\bQ)\Big)\Big(x^\dagger_{i,n}(\bq)\,x^\dagga_{i,n}(\bq) \\
&\qquad\qquad\qquad\qquad\quad +
p_{i,n}^\dagger(\bq-\bQ)\,p^\dagga_{i,n}(\bq-\bQ)\Big)\,,\\
H_{1} &\simeq SJ_1\sum_{n,\bq}\bigg(
\gamma(\bq)\,\mathbf{X}_{1,n}(\bq)\cdot\mathbf{X}_{2,n}(-\bq) + H.c.\bigg)\,,\\
H_{3} &\simeq SJ_3\sum_{n,\bq}\bigg(
\gamma(\bq)\,\mathbf{X}_{1,n}(\bq)\cdot\mathbf{X}_{2,n+1}(-\bq) + H.c.\bigg)\,,
\eal
where $E_0=2N L\,S(S+1)\,J_2(\bQ)$, $\gamma(\bq)=\gamma(\bq,\delta=0)$, and 
\bealn
\mathbf{X}_{i,n}(\bq) = \bd{n}_{1,i,n}\,x_{i,n}(\bq) 
+ \bd{n}_{2,i,n}\,p_{i,n}(\bq-\bQ)\,.
\eal
We note that $H_2$ does not depend on the choice of $\bd{n}_{3,i,n}$, reflecting the classical accidental degeneracy, 
unlike $H_1+H_3$. 
We start treating $H_1$ and $H_3$ within perturbation theory. The unperturbed Hamiltonian 
$H_2$ is diagonalised through the canonical transformation 
\bealn
x^\dagga_{i,n}(\bq) &\to \sqrt{\,K(\bq)\;}\;x^\dagga_{i,n}(\bq)\,,\\ 
p^\dagga_{i,n}(\bq) &\to \sqrt{\,\fract{1}{\,K(\bq)\,}\;}\;p^\dagga_{i,n}(\bq)\,,
\eal
where 
\bealn
K(\bq)^2 &=  \fract{\; J_2(\bnot)-J_2(\bq)\;}{\;J_2(\bnot)+J_2(\bq)\;}\;.
\eal
In the transformed basis, 
\bealn
H_2 &= \fract{1}{2}\sum_{i,n\,\bq}\omega_2(\bq)\,\Big(
p^\dagger_{i,n}(\bq)\,p^\dagga_{i,n}(\bq)+x^\dagger_{i,n}(\bq)\,x^\dagga_{i,n}(\bq)\Big)\,,
\eal
is diagonal, and the spin-wave energy dispersion is  
\bealn
\omega_2(\bq) &= 2S\,\sqrt{\,J_2(\bnot)^2-J_2(\bq)^2\;}\;.
\eal
The free energy in perturbation theory can be written as $F = \sum_\ell\,F_\ell$, where 
$F_\ell$ is of $\ell$-th order in $H_1+H_3$, and $F_0$ is the unperturbed free energy of the Hamiltonian $H_2$. 
Notice that only even-order terms are non vanishing, thus $\ell=0,2,4,\dots$ . 
Given the evolution operator in imaginary time, 
\bealn
S(\beta) &= T_\tau\Bigg( \esp{-\int_0^\beta d\tau\, \big(H_1(\tau)+H_3(\tau)\big)}\,\Bigg)
= \sum_\ell\, S_\ell(\beta)\,,
\eal
where $H_a(\tau)$, $a=1,3$, evolves with the Hamiltonian $H_2$, 
the second order correction to the free energy is readily found to be 
\beal
F_2 &= -T\,\langle\,S_2(\beta)\,\rangle \\
&= - \fract{\Xi_2(T)}{J_2}\,\sum_n\bigg[ 
J_1^2\,\big(\bd{n}_{3,1,n}\cdot\bd{n}_{3,2,n}\big)^2\\
&\qquad\qquad\qquad\qquad  + J_3^2\,\big(\bd{n}_{3,1,n}\cdot\bd{n}_{3,2,n+1}\big)^2
\bigg]\,,
\label{model: F_2}
\eal
where
\bealn
\Xi_2(T) &= J_2\,S^2\,\sum_{\bq}\,T\,\sum_{\lambda}\, 
\big|\gamma(\bq)\big|^2\,K(\bq)^2\\
&\qquad\qquad\qquad  \left(\fract{\omega_2(\bq)}{\;\omega_\lambda^2+\omega_2(\bq)^2\;}\right)^2 >0\,,
\eal
with $\omega_\lambda = 2\pi\lambda T$, $\lambda \in \mathbb{Z}$, bosonic Matsubara frequencies. Even without explicitly evaluating $\Xi_2$, we can 
conclude that the free-energy gain at second order in $H_1+H_3$ is maximised by 
$\bd{n}_{3, 1,n}\cdot\bd{n}_{3, 2,m} = \pm 1$, with $m=n,n+1$, which reduces the classical degeneracy to 
$4^L$ configurations, where $L$ is the total number of layers in the system. Such residual degeneracy is split by a fourth order correction to the free energy 
proportional to $J_1^2\,J_3^2$ that reads 
\beal 
F_4 
&= -\fract{J_1^2\,J_3^2}{J_2^3}\;\Xi_4(T)\,\sum_n\, \big(\bd{n}_{3,1,n}\cdot\bd{n}_{3,2,n}\big)\\
&\qquad \Big[\big(\bd{n}_{3,1,n}\cdot\bd{n}_{3,2,n+1}\big)
+\big(\bd{n}_{3,1,n-1}\cdot\bd{n}_{3,2,n}\big)\Big]\,,
\label{model: F_4}
\eal
where  
\bealn
\Xi_4(T) &=2S^4\,J_2^3\,T\,\sum_\lambda\, \sum_\bq\,
\big|\gamma(\bq)\,\gamma(\bq+\bQ)\big|^2\\
&\qquad\qquad\qquad\qquad 
\fract{\omega_\lambda^2}{\;\big(\omega_\lambda^2+\omega_2(\bq)^2\big)^3\;}
>0\,.
\eal
We remark that, despite $\omega_2(\bq)$ vanishes linearly at $\bq=\bnot,\bQ$, both $\Xi_2(T)$ and $\Xi_4(T)$ 
are non-singular. \\
The fourth order correction $F_4$ in Eq.~\eqn{model: F_4} has a twofold effect: it forces $\bd{n}_{3,1,n}\cdot\bd{n}_{3,2,n}$ 
to be the same on all layers and, in addition, stabilises a ferromagnetic inter-layer stacking. Therefore, the ground state 
manifold at fourth order in $H_1+H_3$ is spanned by $\bd{n}_{3,1,n}=\bd{n}_3$ 
and $\bd{n}_{3,2,n}=\sigma\,\bd{n}_3$, where $\bd{n}_3$ is an arbitrary unit vector reflecting the 
spin $SU(2)$ symmetry, and $\sigma=\pm 1$ is associated with the global $C_4\to C_2$ symmetry breaking. \\

\noindent
Similarly to the single-plane $J_1-J_2$ 
model~\cite{J1-J2-original}, the above results imply that an additional term must be added to the semiclassical spin action. 
Specifically, if we introduce the Ising-like  
fields $\sigma_n(\bR)=\bd{n}_{3,1,n}(\bR)\cdot\bd{n}_{3,2,n}(\bR)$  
and $\sigma_{n+1/2}(\bR)=\bd{n}_{3,1,n}(\bR)\cdot\bd{n}_{3,2,n+1}(\bR)$, Eqs~\eqn{model: F_2} and \eqn{model: F_4} imply that, at the leading orders in $J_1$ and $J_3$, the effective action in the continuum limit includes 
the quadrupolar coupling term~\cite{J1-J2-original} 
\bw
\beal
A_\text{Q}&\simeq -\sum_n\,\int\! d\bR\,\Bigg\{
\fract{\Xi_2(T)}{TJ_2}\Big( J_1^2\,\sigma_n(\bR)^2 +J_3^2\,\sigma_{n+1/2}(\bR)^2\Big)
+\fract{\Xi_4(T)J_1^2 J_3^2}{TJ_2^3}\;\sigma_n(\bR)
\Big(\sigma_{n+1/2}(\bR) + \sigma_{n-1/2}(\bR)\Big)\Bigg\}\,. 
\label{model: Ising model}
\eal
\ew
We expect~\cite{J1-J2-original} an Ising transition to occur at a critical temperature 
$T_c$, below which $\langle\sigma_n(\bR)\rangle = m_1$,  
$\langle\sigma_{n+1/2}(\bR)\rangle = m_3$, with $m_1\,m_3>0$. However, this transition is now three-dimensional and brings along the AFS magnetic ordering, thus a finite N\'eel temperature $T_N\simeq T_c$. \\
To get a rough estimate of $T_N$, we assume that, upon integrating out the spin degrees of freedom, the classical action describes an anisotropic three-dimensional ferromagnetic Ising model with exchange constants $I_1$ on layers $n$, $I_3$ on layers $n+1/2$, and $I_\perp<I_1$ between layers. Hereafter, we take for simplicity $J_1=J_3$, thus $I_1=I_3\equiv I_\parallel$.
We then note that, if we use the above estimates of $J_2\simeq 9.3~\text{meV}$ and $J_{1}= J_3 \simeq -1.17~\text{meV}$, 
the 2D Ising critical temperature with $S=3/2$ of each layer $n$ and $n+1/2$ is about 
$0.4\, (S+1/2)^2 J_2\simeq 173~\text{K}$~\cite{Capriotti-PRL2004}. That corresponds to $I_\parallel\simeq 6.6~\text{meV}$~\cite{Yurishchev-2004}. The 3D Ising critical temperature $T_c\simeq T_N$ depends on the anisotropy ratio 
$I_\perp/I_\parallel$~\cite{Yurishchev-2004}, being at most 
$345~\text{K}$ for $I_\perp/I_1\simeq1$, and reaching the observed 
$T_N\simeq 290$~\cite{Abushammala2023} 
at $I_\perp/I_1\simeq 0.75$, which we find not unrealistic. 

\noindent
However, BaCoS$_2$ remains orthorhombic above $T_N$, which implies that 
the structural $C_4\to C_2$ symmetry breaking occurs earlier than magnetic 
ordering. Therefore, even though the effects of the orthorhombic distortion 
on the electronic structure are rather small, see Table~\ref{tab:Hops}, 
it is worth repeating the above discussion assuming from the start that 
$\sigma_n(\bR)=\bd{n}_{3,1,n}(\bR)\cdot\bd{n}_{3,2,n}(\bR) = 1$ (AFS-a), so that 
\bealn
\sigma_{n+1/2}(\bR)&=\bd{n}_{3,1,n}(\bR)\cdot\bd{n}_{3,2,n+1}(\bR)\\
&=\bd{n}_{3,1,n}(\bR)\cdot\bd{n}_{3,1,n+1}(\bR)\,.
\eal 
It follows that the quadrupolar term~\eqn{model: Ising model} becomes 
\bea
A_\text{Q}&\simeq& -\sum_n \int\! d\bR\, \bigg\{
K(T)\,\Big(\bd{n}_{3,1,n}(\bR)\cdot\bd{n}_{3,1,n+1}(\bR)\Big)^2\nonumber\\
&&\qquad \qquad \quad  
+h(T)\,
\bd{n}_{3,1,n}(\bR)\cdot\bd{n}_{3,1,n+1}(\bR) \bigg\}\,,
\label{model: Ising model-2}
\eea
with $K(T)\gg h(T)>0$, and provides a direct coupling between adjacent planes able to stabilise the 3D magnetic order. For $h(T)=0$, C-AFS, $\bd{n}_{3,1,n}\cdot\bd{n}_{3,1,n+1}=1$, and G-AFS,
$\bd{n}_{3,1,n}\cdot\bd{n}_{3,1,n+1}=-1$, 
are degenerate. In this case, the N\'eel transition occurs simultaneously with 
the Ising transition that makes the system choose either C-AFS or G-AFS. 
Moreover, since the two types of stacking are separated by a macroscopic energy barrier, $T_N$ can be large despite a weak 
spin-wave dispersion along the $c$-axis, not in disagreement with 
neutron scattering experiments~\cite{Shamoto-JPSJ1997,Shamoto-PRR2021}. 
The finite $h(T)$, which DFT+U predicts to be only few kelvins at $T=0$, 
eventually favours C-type stacking. Nonetheless, we expect that the N\'eel transition should still maintain a strong Ising-like character. 



\section{Conclusions}
We showed by \textit{ab initio} calculations that the low-temperature phase of BaCoS$_2$ critically depends on the specific Co $d$-orbitals where the three unpaired electrons lie. Indeed, the directionality of those orbitals and the 
significant hybridisation with ligand sulfur atoms are ultimately responsible of the 
magnetic frustration implied by the columnar magnetic order despite the lattice being bipartite. Moreover, the $d_{xz}-d_{yz}$ doublet hosting a single hole 
entails orbital degeneracy in tetragonal symmetry that has to be lifted at low temperature. However, contrary to the expectation that the orthorhombic distortion 
should make the hole choose either $d_{xz}$ or $d_{yz}$, we find that the lowest-energy state has an antiferro-orbital order, thus both orbitals almost equally occupied on average, which breaks the non-symmorphic symmetry and conspires with magnetism to stabilise an insulating phase. For the same reason, the orthorhombic distortion that persists above the N\'eel transition temperature~\cite{Mandrus1997,Abushammala2023} does not produce the large   
nematicity that could efficiently remove magnetic frustration 
should the occupancies of $d_{xz}$ and $d_{yz}$ be very different. \\
Magnetic frustration and orbital degeneracy are also responsible of the multitude of states that we find within 20~meV above the lowest energy one, which corresponds to planes with columnar spin-order and antiferro-orbital order stacked ferromagnetically, both in spin and orbital, along the $c$-axis. 
This abundance of low-energy excitations revealed by GGA+U already hints at the order-from-disorder origin of the magnetic transition that we uncover by studying the effective Heisenberg model built from the Wannierization of the band structure, and which represents a three-dimensional generalisation of the $J_1-J_2$ Heisenberg model on a square lattice.
Similarly to the latter~\cite{J1-J2-original}, the model in Fig.~\ref{J1-J2-model} 
displays a finite-temperature order-from-disorder Ising-like transition, which is now three-dimensional and thus able to drive also a finite temperature N\'eel transition. This result rationalises in simple terms some of the puzzling features of 
BaCoS$_2$, specifically, the strong Ising-like character of the N\'eel transition~\cite{Mandrus1997,Abushammala2023}, the large $T_N$ despite the 
weak $c$-axis dispersion of the magnetic excitations~\cite{Shamoto-JPSJ1997,Shamoto-PRR2021}, and the anomalous electronic properties 
above $T_N$~\cite{SantosCottin2018} hinting at the presence of strong low-energy fluctuations.\\

\noindent
We finally draw a parallel to another class of materials where orbital, magnetic and structural order parameters are intertwined and play an important role, namely iron pnictide superconductors (FeSC)~\cite{Fernandes2014} and FeSe ~\cite{glasbrenner2015,baek2015, Wang2015, Boehmer2016}. 
In many (mainly electron-doped) iron-based superconductors the disordered phase at high temperature first spontaneously breaks $C_4$ symmetry when cooling below a critical temperature $T_{\mathrm{nem}}$, thus entering a nematic phase~\cite{Fernandes2014}.
Only at a lower temperature $T_N<T_{\mathrm{nem}}$, 
also spin $SU(2)$ is broken and a stripe-ordered magnetic long-range order emerges~\cite{Fernandes2014, Wang2015}. The N\'eel temperature can be quite large 
as in \BCS\, or even vanishing within experimental accuracy as in FeSe, 
where $T_N\not=0$ is observed only under pressure~\cite{kothapalli2016}.\\
Also from a model point of view, our description of \BCS\ fits into the modelling of these materials. Indeed, the key role played by several order parameters in 
\BCS\ has parallels to the phenomenological Landau free energy description of FeSC~\cite{Fernandes2014}. 
Besides itinerant multi-orbital Hubbard models \cite{Chubukov2008,Stanev2008,Graser2009}, also spin-1 Heisenberg models conceptually similar to ours have been proposed for iron-based superconductors and FeSe~\cite{Hu2012,glasbrenner2015}.
There, however, instead of achieving the $C_4$ symmetry breaking via an order-from-disorder phenomenon, that is often assumed from the start~\cite{zhao_2009} or by explicitly adding bi-quadratic spin exchanges~\cite{Hu2012} 
that mimic the quadrupolar terms \eqn{model: Ising model}.
\\
Moreover, our \textit{ab initio} simulations for \BCS\ predict that the lowest-energy phase has not ferro-orbital order, as often discussed in the context of FeSC, but  rather an anti-ferro orbital ordering that breaks the non-symmorphic symmetry instead of   $C_4$. Therefore,
\BCS\ seems to realise a situation where collinear magnetism and orthorhombicity do not imply orbital nematicity, unless for specific crystal structures under pressure.\\
\BCS\ could therefore add yet another piece to the puzzle of understanding the intricate interplay of structural, magnetic and electronic degrees of freedom in strongly correlated materials.

\acknowledgments
We are thankful to H. Abushammala, A. Gauzzi, and Y. Klein for fruitful discussions. 
We acknowledge the allocation for computer resources by the French Grand \'Equipement National de Calcul Intensif (GENCI) under the project numbers A0010906493 and A0110912043.
M.F. acknowledges financial support from the European Research Council (ERC),
under the European Union's Horizon 2020 research and
innovation programme, Grant agreement No. 692670 "FIRSTORM".
\bibliography{Biblio}
\end{document}